# Bulk photovoltaic effect in BaTiO$_3$-based ferroelectric oxides: An experimental and theoretical study

Subhajit Pal[1], S. Muthukrishnan[1], Banasree Sadhukhan[2], Sarath N. V.[1], D. Murali[3], and Pattukkannu Murugavel[1*]

[1]*Department of Physics, Indian Institute of Technology Madras, Chennai 600036, India*
[2]*Leibniz Institute for Solid State and Materials Research IFW Dresden, Helmholtzstr. 20, 01069 Dresden, Germany*
[3]*Department of Science, Indian Institute of Information Technology Design and Manufacturing, Kurnool, Andhra Pradesh, 518002, India*

## Abstract

The bulk photovoltaic effect exhibited by the non-centrosymmetric system gains research interest due to the observed large open-circuit voltage. The ferroelectric systems exhibiting anomalous photovoltaic effect are mostly crystallized with multiphase coexistence. Hence, the computational difficulty in building a multi-phase system restricts the detailed photovoltaic studies through phenomenological and shift current theory. In this work, ferroelectric Ba$_{1-x}$(Bi$_{0.5}$K$_{0.5}$)$_x$TiO$_3$ (BBKT) oxide is designed to crystallize in single-phase tetragonal symmetry with improved polarization characteristics, and it is found to exhibits large PV response. Both experimental and theoretical studies on BBKT samples reveal ~18% reduction in bandgap compared to the parent BaTiO$_3$. Short-circuit current measured as a function of light intensity and light polarization angle reveal linear and sinusoidal response, respectively. The observed features are in accordance with phenomenological theory. Remarkably, $x = 0.125$ sample displays ~8 times higher open-circuit voltage (7.39 V) than the parent compound. The enhanced PV effect is attributed to the large shift current along $z$-direction as evidenced from the additional charge-center shift of valence band occupied by O-2$p$ orbital and conduction band occupied by Bi-6$p$ orbital. Notably, the degenerate Bi-$p_z$ state at conduction band minimum in BBKT favours the large shift current response in the $z$-direction.

*muruga@iitm.ac.in

# I. INTRODUCTION

The anomalous photovoltaic (PV) phenomenon observed in the ferroelectric system is an emerging research field in the context of the next-generation PV materials.[1-4] In particular, the reported giant photogenerated field (~ $10^4$ V/cm) in Fe doped LiNbO$_3$ single crystal and the demonstrated power-conversion efficiency exceeding the Shockley-Queisser limit at nano-scale in BaTiO$_3$ (BTO) thin film, make the ferroelectric systems as the promising candidate for PV studies.[5,6] Several mechanisms are proposed to explain the anomalous ferroelectric PV response.[2,7-11] Among them, the bulk photovoltaic effect (BPVE), associated with the violation of the detailed balance principle, is a widely accepted mechanism.[2,6,10,11]

The anomalous PV effects in several non-centrosymmetric systems having single and multiple phase co-existence such as BTO, PbTiO$_3$, BiFeO$_3$, KNbO$_3$, (K,Ba)(Ni,Nb)O$_{3-\delta}$, ZnSnP$_2$, etc., are analyzed either through phenomenological approach or by shift current theory.[12-18] In the phenomenological model, the PV current consists of linear and circular components as expressed by,

$$J_i^{pv} = \beta_{ijk} e_j e_k^* I_o + \mathrm{i}\beta_{ik}[ee^*]_k I_o \qquad (1)$$

Where, $I_o$ is the light intensity, $e_j$ and $e_k^*$ are components of the incident light polarization vector. $\beta_{ijk}$ and $\beta_{ik}$ are third and second rank tensors, respectively.[10,11] The linear component in the first term is the current response under linearly polarized light in pyro or piezoelectric crystals. The circular component in the second term is the current response under circularly polarized light in gyrotropic crystals.[2] In the shift current model, the photocurrent originates from the shift in charge-center in real space due to the difference in the Berry connection between the valence and conduction bands involved in the optical excitation process.[14,15,19,20]

The large linear BPVE demonstrated on (001) BTO thin film is correlated to sample dimension and characteristic length by calculating the tensor element $\beta_{31}$ through phenomenological approach.[10] On the other hand, the enhanced PV effect in (K,Ba)(Ni,Nb)O$_3$- $_\delta$, 0.5Ba(Zr$_{0.2}$Ti$_{0.8}$)O$_3$-0.5(Ba$_{0.7}$Ca$_{0.3}$)TiO$_3$, Ba$_{1-x}$(Bi$_{0.5}$Li$_{0.5}$)$_x$TiO$_3$, and Ba(Ti$_{1-x}$Sn$_x$)O$_3$ systems are attributed to the increase in shift current due to orbital mixing at the conduction band-edge and the structural preference.[14,16,21,22] Also, the spectral dependence studies of the shift current response on the ZnSnP$_2$ system is found to be originated from the real space charge-center shift between valence and conduction electrons of P-3$p$ and Sn-5$p$ orbitals.[15] However, to get detailed insight on the PV characteristics having a strong correlation with the bulk PV tensor elements and electronic band structure, the analysis based on both the phenomenological model and shift current mechanism is necessary.[2] Nevertheless, such studies are not reported on the ferroelectric PV systems.

In the present work, the PV response is investigated on the modified BTO system, Ba$_{1-x}$(Bi$_{0.5}$K$_{0.5}$)$_x$TiO$_3$ (BBKT) ($x$ = 0.0, 0.05, 0.075, 0.1, 0.125, and 0.15), and the results are analysed through phenomenological approach and shift current theory. The BBKT system is designed by doping Bi$^{3+}$ at the Ba$^{2+}$ site to capitalize its polarization characteristic due to 6$s^2$ lone pair contribution.[21,23] A monovalent K$^{1+}$ is chosen as a co-dopant for charge compensation. For the ease of theoretical analysis on the BPVE, the sample with a single-phase is preferable. In this regard, K$^{1+}$ with $r_K$ = 1.51 Å could be an appropriate dopant that could reduce the lattice distortion and minimize the structural inhomogeneities compared to similar dopants like Li$^{1+}$, which has smaller ionic radii.[24] Consequently, the fabricated BBKT samples revealed single-phase characteristics, and the observed PV response is found to be in accordance with the phenomenological theory. It is observed that the shift current response in the BBKT system has additional contribution from real space charge-center shift of the conduction band occupied by Bi-6$p$ orbital. Also, the analysis of the spectral-dependence of the photocurrent response through shift current mechanism strongly suggests the superior PV response of BBKT samples over the parent BTO system at higher incident photon energy range. The combined experimental and theoretical studies on the BPVE are presented in this work.

## II. METHODS: EXPERIMENT AND COMPUTATION

### A. Experimental details

$Ba_{1-x}(Bi_{0.5}K_{0.5})_xTiO_3$ ($x$ = 0.0, 0.05, 0.075, 0.1, 0.125, and 0.15) were fabricated by solid-state synthesis method using $BaCO_3$ (purity ≥ 99.9%), $TiO_2$ (purity ≥ 99.9%), $Bi_2O_3$ (purity ≥ 99.9%), and $K_2CO_3$ (purity ≥ 99.9%) as precursors. The stoichiometric quantities of the precursors were ground and calcined at 700 ºC for 4 h followed by 800 ºC for 6 h. The pellets of 12 mm diameter made from the calcined powder of various compositions were subjected to heat treatment at optimized temperatures (1300 to 1100 ºC) for 4 h. The X-ray diffraction (XRD) experiments were performed on the powdered samples for phase confirmation using PANanalytical X-ray diffractometer. The Rietveld refinements were carried out on the XRD patterns by FULLPROF software.[25] Dielectric and polarization measurements were carried out on Ag coated 12 mm diameter pellets. Temperature-dependent dielectric permittivity studies were done from NovoControl impedance analyzer. Polarization measurements were carried out by the Radiant Technology loop tracer. The Ultraviolet-Visible–Near Infrared (UV–VIS–NIR) Spectro-Photometer (Perkin Elmer LAMBDA 950) was employed for optical bandgap measurements. The samples of 0.25 mm thickness with thermally evaporated 0.5 mm diameter of gold top electrodes and Ag bottom electrodes were used for PV measurements. Before PV measurements, the samples were subjected to poling at +30 kV/cm electric field for 1 h followed by shorting the electrodes for 48 h to remove the parasitic surface charges. PV measurements were carried out using 405 nm diode laser (MDL-III-405) as a light source and Keithley electrometer (6517B) as a measuring unit. Light polarization angle-dependent PV studies were performed using Holmarc linear film polarizer, having 99% polarizing efficiency.

### B. Computational methods

The density-functional theory (DFT) calculations were performed with the localized atomic orbital basis as implemented in the full-potential local-orbital (FPLO) code.[26] The exchange-correlation functions were considered at the generalized gradient approximation (GGA) level with k-grids 8 × 8 × 8.[27] The total ionic charges are calculated using the Bader Atom Molecule approach. The momentum matrix elements for the optical conductivity were calculated within the linear response as implemented in the FPLO. The optical current density $J_i$ generated by the electric field $E_j$ is given by, $J_i = \sigma_{ij}E_j$. Complex dielectric function $\varepsilon(\omega) = \varepsilon_1(\omega) + i\varepsilon_2(\omega)$ describes the behavior of materials under incident light, where $\varepsilon_1(\omega)$ and $\varepsilon_2(\omega)$ are real and imaginary parts of the dielectric function, respectively. The imaginary part $\varepsilon_2(\omega)$ is obtained from the electronic structure through the joint density of the states and the momentum matrix elements between occupied and unoccupied states.

$$\varepsilon_2^{ij}(\omega) = \text{Im}[\varepsilon_{ij}(\omega)] = -\frac{4\pi^2 e^2}{m_0^2 \omega^2} \int d\mathbf{k} \sum_{n,l} (f_n - f_l) \times \frac{\langle k_n|\hat{v}_i|k_l\rangle\langle k_l|\hat{v}_j|k_m\rangle}{(E_{kn} - E_{kl} + \hbar\omega - i\delta)}$$

Where, $i; j = (x; y)$ are the Cartesian coordinates, $p$ is the momentum operator, $k_n$ is the eigenfunction with eigenvalue $E_{k_n}$, and $f(k_n)$ is the Fermi distribution function. The optical conductivity $\sigma_{ij}(\omega)$ is directly related to dielectric function and can be expressed as $\sigma_{ij}(\omega) = \frac{\omega \varepsilon_2^{ij}(\omega)}{4\pi}$.

To calculate the shift current conductivity, the tight-binding Hamiltonian is constructed by projecting the wave function in atomic basis into a reduced symmetric atomic orbital like Wannier functions with Ba-6$s$, 5$p$, 5$d$, Ti-3$d$, O-2$s$, and 2$p$ orbitals for BTO in energy range (-4.9, 14.7) eV. Bi-4$f$, 5$d$, 6$s$, 6$p$, and K-4$s$ orbitals are also taken into account for BBKT structures. We used multi-band approach to calculate the shift current conductivity and is given by,[28-30]

$$\sigma_{ij}^k = \frac{|e|^3}{8\pi^3 \omega^2} Re\left\{\varphi_{ij} \sum_{\Omega=\pm\omega} \sum_{l,m,n} \int_{BZ} dk(f_l - f_n) \times \frac{\langle k_n|\hat{v}_i|k_l\rangle\langle k_l|\hat{v}_j|k_m\rangle\langle k_m|\hat{v}_k|k\rangle}{(E_{kn}-E_{km}-i\delta)(E_{kn}-E_{kl}+\hbar\Omega-i\delta)}\right\} \quad (2)$$

The real part of the integral describes the shift current response under linearly polarized light. The conductivity $\sigma_{ij}^k$ ($i; j; k = x; y; z$) is a third rank tensor representing the photocurrent density $J_k$ generated by an electrical field via $J_k = \sigma_{ij}^k E_i^* E_j$. $\varphi_{ij}$ is the phase difference between the driving field $E_i^*$ and $E_j$. The integral in Eq. (2) describes the shift current response under linearly polarized light.

## III. RESULTS AND DISCUSSION

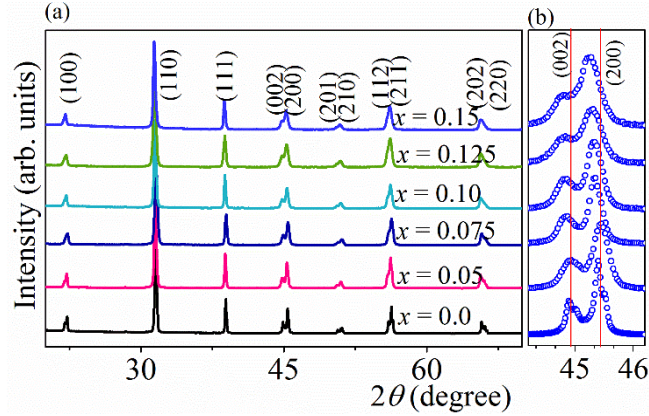

FIG. 1. (a) The XRD patterns of $x$ = 0.0, 0.05, 0.075, 0.1, 0.125 and 0.15 BBKT samples. (b) Enlarged XRD pattern around (002)/(200) diffraction peaks. The solid lines are drawn for the reference with respect to the parent sample.

To remove the mechanical stress, if any, arises during the grinding process, the powder obtained from the sintered pellets are subjected to annealing at 200 ºC for 4 h. The XRD patterns of the BBKT samples with $x$ = 0.0, 0.05, 0.075, 0.1, 0.125, and 0.15 are recorded on the stress free annealed powder and the respective plots are displayed in Fig. 1. The patterns indicate that the synthesized samples are free from impurity phases. From the initial evaluation of the XRD data, the patterns are index to the tetragonal crystal symmetry having *P*4*mm* space group. The evolution of the Bragg's peaks (002) and (200), which are signature peaks for the tetragonal crystal structure, are also plotted in Fig. 1(b) for reference. Keeping in mind the fact about multi-phase coexistence in parent BTO system, the single and double phase models are considered to refine the obtained XRD patterns.[24,31] Consequently, the combination of tetragonal (*P*4*mm*) with orthorhombic (*Amm*2) and monoclinic (*Pm*) phase models are selected.

To determine the possible crystal symmetry of the BBKT samples, the combination of tetragonal (*P4mm*) with orthorhombic (*Amm*2) and monoclinic (*Pm*) phase models are chosen for the Rietveld refinement. The *Amm*2 and *Pm* are considered in two-phase models along with *P4mm* phase due to the recent observation of the subtle monoclinic phase together with the orthorhombic phase in the tetragonal BTO system.[24,31,32] The refined patterns near (002) and (200) Bragg's peaks of all the phase models are displayed in Fig. 2. The figure depicts that the satisfactory fitting is obtained, when the BBLT samples are refined with *P4mm* phase. In contrast, the double phase models, *P4mm*+*Amm*2, and *P4mm*+*Pm*, are resulted in unsatisfactory fitting parameters with poor fitting compared to *P4mm* phase model. It is to be noted that, unlike $Li^{1+}$ doped sample, the refinements reveal single-phase characteristics of the BBKT samples.[24] The refinement parameters shown in TABLE I indicate the increase in the lattice parameters (*a* and *c*) with composition. This is further evidenced from the shift in (200) and (002) Bragg peaks towards a lower diffraction angle, as shown in Fig. 1(b). These shifts are in accordance with decrease in average ionic radii of the dopant ions ($r_{Bi}$ = 1.17 Å, and $r_K$ = 1.51 Å) in comparison to the parent ion $Ba^{2+}$ ($r_{Ba}$ = 1.42 Å).[33,34]

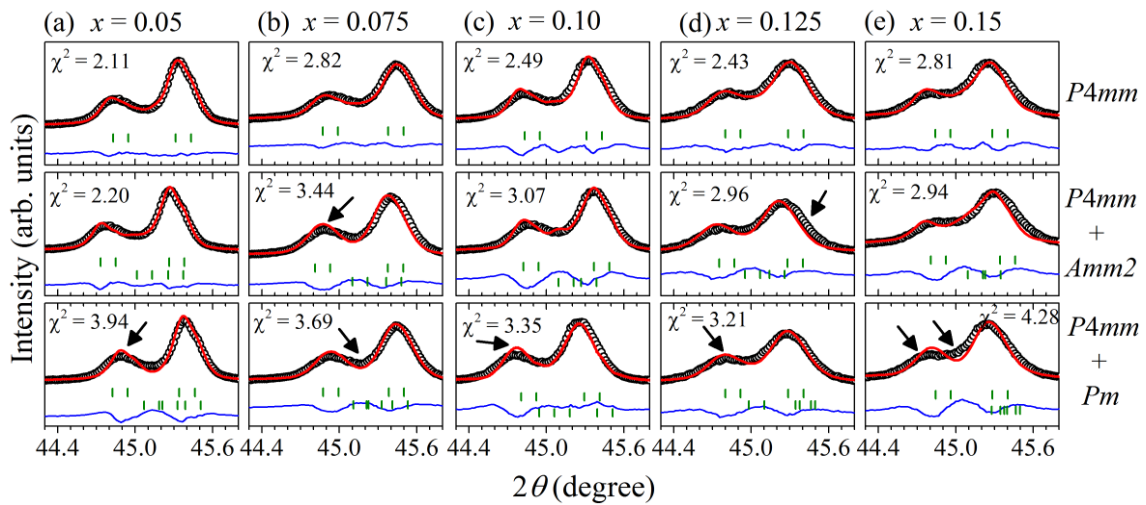

FIG. 2. XRD patterns represent Bragg profiles of (002) and (200) for the BBKT system fitted *P4mm*, *P4mm*+*Amm*2, and *P4mm*+*Pm* phase models for (a) *x* = 0.05, (b) *x* = 0.075, (c) *x* = 0.10, (d) *x* = 0.125, and (e) *x* = 0.15 samples. Open symbol, solid line, and vertical line represent

experimental points, refined data, and Bragg profiles, respectively. The bottom solid lines represent the error between experimental and calculated data points.

TABLE I. The refinement parameters for $x$ = 0.0, 0.05, 0.075, 0.10, 0.125, and 0.15 BBKT samples.

| $x$ | $\chi^2$ | $R_{wp}$ (%) | $a$ (Å) | $c$ (Å) |
|---|---|---|---|---|
| 0.0 | 3.14 | 8.76 | 3.993 | 4.003 |
| 0.05 | 2.11 | 9.15 | 3.996 | 4.037 |
| 0.075 | 2.82 | 10.2 | 3.998 | 4.04 |
| 0.10 | 2.49 | 9.85 | 3.999 | 4.043 |
| 0.125 | 2.43 | 9.32 | 4.001 | 4.047 |
| 0.15 | 2.81 | 10.1 | 4.003 | 4.048 |

To understand the ferroelectric feature of the BBKT samples, polarization (P) versus electric field (E) measured at 4 Hz are shown in Fig. 3(a). The figure reveals typical P-E hysteresis loops of the ferroelectric system but with significant remnant polarization ($P_r$) variations with compositions. The $P_r$ values of $x$ = 0.0, 0.05, 0.075, 0.1, 0.125 and 0.15 samples are 5.2, 7.93, 8.29, 6.84, 6.23 and 4.45 μC/cm$^2$, respectively. Although the observed initial increment in $P_r$ value up to $x$ = 0.075 could be correlated to the Bi-6$s^2$ lone pair contribution, the decreasing trend for higher doping concentrations could be attributed to the dopant's adverse effect on Ti-O hybridization.[21] To understand the ferroelectric phase transition, the temperature-dependent real part of the dielectric permittivities ($\varepsilon_1$) measured at 10 kHz for all samples are given in Fig. 3(b). The parent compound BTO depicts two peaks at 10 and 124 °C corresponding to orthorhombic to tetragonal and tetragonal to cubic phase transitions.[32] On the other hand, the doped samples reveal only tetragonal to cubic phase transition ($T_C$) within the measurement range. With increase in Bi and K concentrations, the $T_C$ is shifting towards higher temperatures from 124 to 197 °C. The phase transition appears with peak broadening associated with diffuse-type phase transition and a decrease in $\varepsilon_1$.[35] These observed trends in $\varepsilon_1$ data could be attributed to increased local charge disorder due to decreased average ionic radius at Ba-site.[24,35]

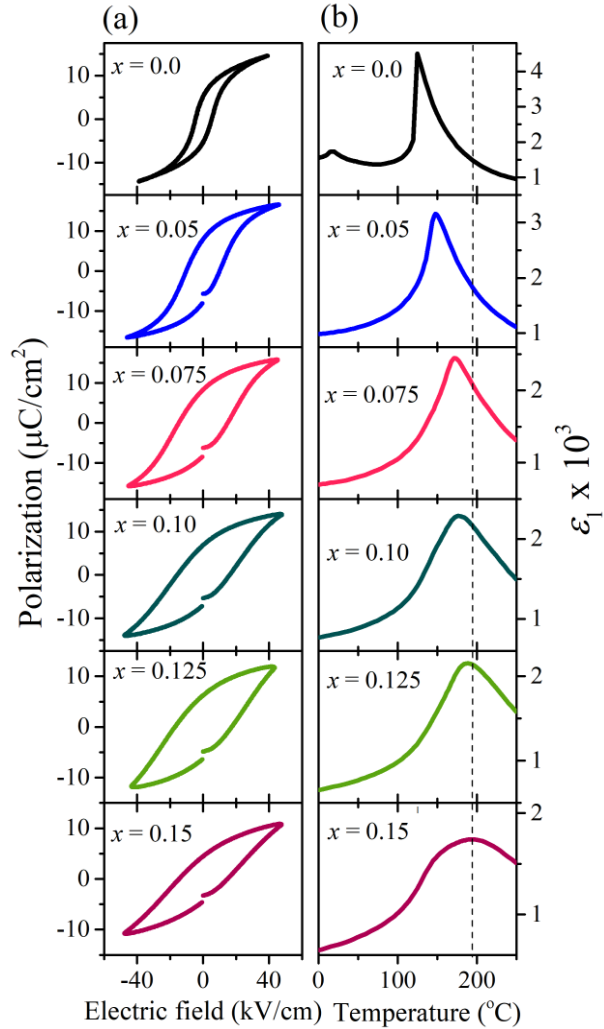

FIG. 3. (a) P-E hysteresis loops and (b) temperature-dependent real parts of the permittivity ($\varepsilon_1$) for $x$ = 0.0, 0.05, 0.075, 0.10, 0.125, and 0.15 samples.

The bandgaps for all the compositions are extracted from the Kubelka-Munk (K-M) plot using the diffused reflectance spectra obtained on the samples.[36] The resultant $[F(R)h\nu]^2$ versus $h\nu$ plots are shown in Fig. 4(a), where $F(R) = \frac{(1-R)^2}{2R}$ is the K-M function, R is the reflectance, $h$ is the Planck's constant, and $\nu$ is the frequency.[36] Interestingly, the first slope in K-M plot at lower energies observed in doped samples yields the direct bandgap and the second transition at 3.2 eV, similar to the undoped BTO sample.[37-39] The bandgap obtained by fitting the K-M plot is displayed as a function of composition in Fig. 4(b). The figure shows that the bandgap is decreased from 3.2 to 2.60 eV (~18%) upon doping and remains nearly the same with further increase in $x$.

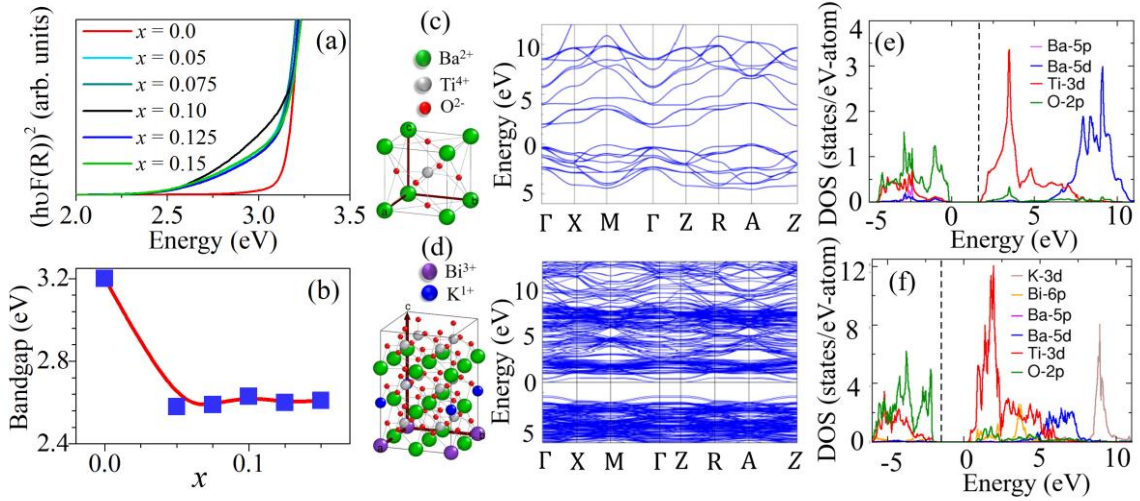

FIG. 4. (a) The K-M plots of BBKT samples and (b) the corresponding bandgap variation with compositions. (c) Super cell structure and band structure of BTO. (d) Super cell structure and band structure of BBKT. DOS of (e) BTO-unit cell and (f) BBKT-223 super cell.

To understand the bandgap variation upon doping in the BTO system, electronic band-structure calculations are carried out on $x = 0.0$ and $0.15$ compositions using unit cell and $2 \times 2 \times 3$ super cell configuration, respectively, by DFT method. Although the multi-phase coexistence in BTO system is reported, for ease of computational framework, the band-structure calculations of the BTO unit cell and doped BBKT sample are performed by considering single tetragonal symmetry with *P4mm* space group. The super cell structures adopted for the calculations are displayed in Fig. 4(c) and 4(d) along with the corresponding band-structures for $x = 0.0$ and $0.15$ samples, respectively. The conduction band minimum (CBM) and valence band maximum (VBM) are observed at Γ and A high symmetry points with the corresponding bandgap of 1.64 eV for the BTO sample, which is ~48% of the experimental value.[40] Whereas, $x = 0.15$ sample displays that both CBM and VBM points are observed at Γ with the corresponding bandgap of 1.35 eV, which is again ~48% of the experimental value. Although, the hybrid functional calculation could estimate the experimental bandgap of the systems, keeping in view of the high computational cost involved in the hybrid functional approach, the GGA level calculation is performed in this work without the loss of physics behind it. However, the reduction in the BBKT sample's bandgap with respect to the parent BTO obtained from the electronic structure calculation (~18%) is in accordance with the experimental observation. The Fig. 4(e) and 4(f) represent the density of states (DOS) plots calculated for $x = 0.0$ and 0.15 samples. In BTO system, the valence band is mostly contributed by O-$2p$ states, whereas the conduction band is contributed by Ti-$3d$ states, as shown in Fig. 4(e). In BBKT, the additional contribution from Bi-$6p$ states to the conduction band and its interaction with Ti-$3d$ state pushes the CBM towards lower energy, which results in bandgap reduction, as shown in Fig. 4(f).

To examine the PV response, the current density ($J$) versus bias voltage (V) is measured in capacitor geometry with ~30 nm thick and ~0.5 mm diameter gold as a top electrode on BBKT samples. The schematic diagram of the experimental set-up is presented in Fig. 5(a). The $J$-V characteristics under dark and 405 nm light illumination (11.9 mW/mm$^2$ intensity) for $x = 0.05, 0.075, 0.1, 0.125$ and 0.15 samples, are displayed in Fig. 5(b). For clarity, the respective plot of the BTO sample is shown in the inset of Fig. 5(b). The figure indicates the BBKT samples display a better PV response with remarkable enhancement in $V_{OC}$ compared to the parent compound. The maximum $V_{OC}$ of 7.39 V ($E_{OC}$ = 296 V/cm) is displayed by $x = 0.125$ sample, which is ~8 times higher in magnitude than the parent BTO. Importantly, the value is 2 times higher than the reported single-crystalline BTO (152 V/cm) under 405 nm illumination[41] and comparable in magnitude with the other polycrystalline ferroelectric systems.[7,8,16,22,42,43]

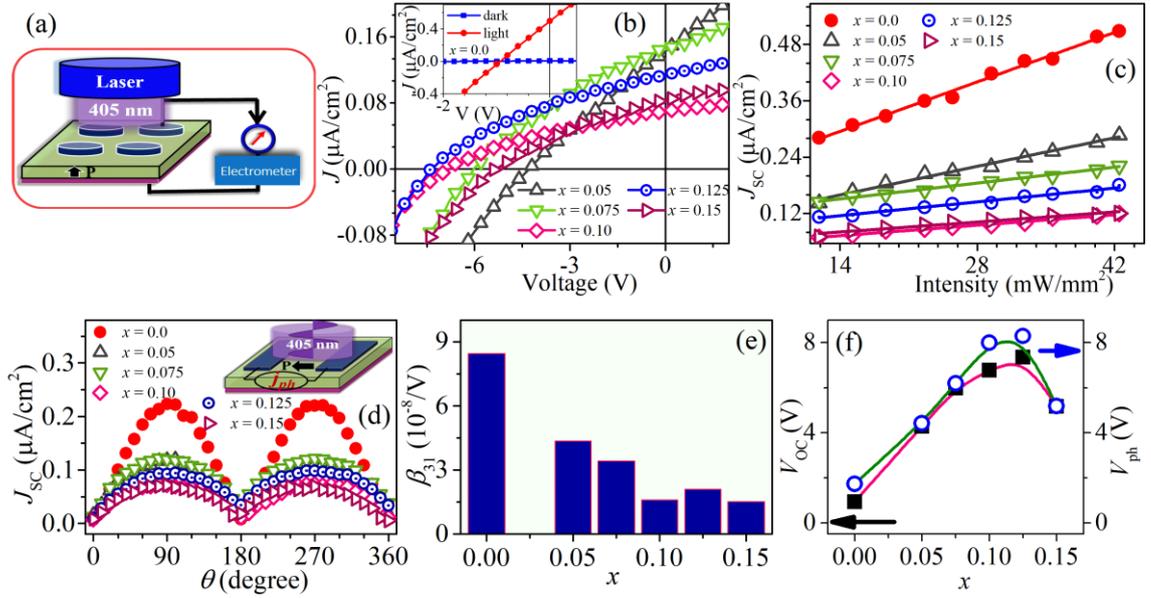

FIG. 5. (a) Schematic of experimental set-up used for PV measurements. (b) $J$-$V$ characteristic under 405 nm illumination measured at 11.9 mW/mm² light intensity for $x$ = 0.05 to 0.15 samples. The inset shows the corresponding plot for $x$ = 0 sample. The $J_{SC}$ plotted as a function of (c) light intensities and (d) light polarization angle for BBKT samples. The inset shows the schematic of the measurement geometry. (e) Extracted $\beta_{31}$ for $x$ = 0.0, 0.05, 0.075, 0.10, 0.125, and 0.15 samples. (f) The composition-dependent experimentally observed $V_{OC}$ and calculated $v_{ph}$ from the phenomenological model.

To probe the BPVE phenomenon, light intensity-dependent $J_{SC}$ is measured and plotted in Fig. 5(c). The Fig. 5(c) shows the linear light intensity-dependence of $J_{SC}$, which is in agreement with BPVE expressed in the first term of Eq. (1).[10,11] To probe it further, the light polarization direction-dependent PV measurements are carried out on the samples. The schematic representation of the experimental geometry is displayed as an inset in Fig. 5(d). Note that the $J_{SC}$ is measured along the in-plane direction as a function of light polarization angle ($\theta$), where $\theta$ is the angle between measured photocurrent direction and linear polarizer axis.[13,18] The obtained $J_{SC}$ versus $\theta$ data is plotted in Fig. 5(d). The measured $J_{SC}$ shows the minimum and maximum values when the light polarization direction is parallel ($\theta = 0°$) and perpendicular ($\theta = 90°$) to the direction of the measured photocurrent, respectively. Importantly, $J_{SC}$'s sinusoidal dependence with the light polarization direction reaffirms the evidence of BPVE in the BBKT samples.[13,18] To examine it further, the wavelength dependent photovoltaic measurement is performed on $x = 0.125$ sample at four different wavelength range using the Xenon Arc Lamp source and the relevant filter assembly. The action spectrum which is the photocurrent response with respect to the wavelength shown in Fig.S1. Nearly negligible value of photocurrent response observed above 405 nm light illumination conditions (490 and 532 nm) explain that the photocurrent response in the BBKT sample is the consequence of shift current effect.

The photocurrent measurement performed under linearly polarized light can give the tensor component $\beta_{31}$ as per Eq. (1). Extracted $\beta_{31}$ from the $J_{SC}$ versus $I_0$ curves plotted in Fig. 5(c) for all samples are presented as histogram in Fig. 5(e). Notably, the highest $J_{SC}$ exhibited by BTO is the consequence of higher $\beta_{31}$ (8.45 × 10$^{-8}$ /V) value.[2,10] As per the phenomenological theory of BPVE, the obtained photovoltage ($v_{ph}$) can be correlated to the steady-state current density ($J_{ph}$) along with dark conductivity ($\sigma_d$) and photo-conductivity ($\sigma_{ph}$) through the expression,

$$v_{ph} = \frac{J_{ph}}{\sigma_d + \sigma_{ph}} d \qquad (3)$$

Where, $d$ is the distance between electrodes.[2] Accordingly, the calculated $v_{ph}$ for the samples using Eq. (3) are shown in Fig. 5(f) along with the obtained $V_{OC}$. Figure 5(f) reveals similar composition-dependent variations of the calculated $v_{ph}$ and observed $V_{OC}$, which gives further experimental evidence for the BPVE.

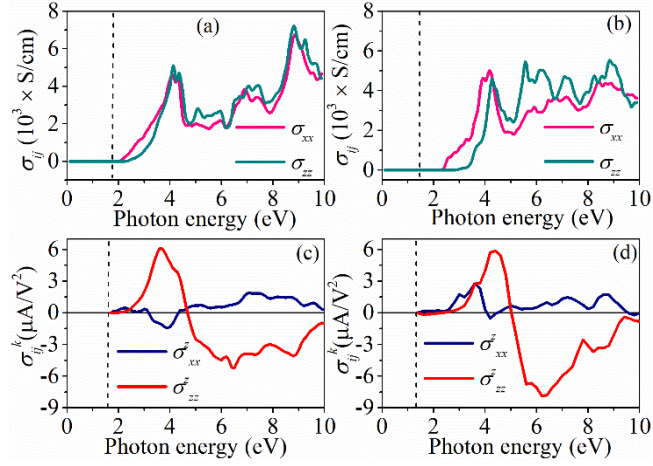

FIG. 6. The optical conductivities $\sigma_{xx}$ and $\sigma_{zz}$ plotted for (a) BTO and (b) BBKT. The shift current conductivities $\sigma^z_{xx}$ and $\sigma^z_{ZZ}$ of (c)BTO and (d) BBKT ($x = 0.15$) system.

To analyze the BPVE from the shift current theory, the components of optical conductivity $\sigma_{xx}$ and $\sigma_{zz}$ calculated from the dielectric function are plotted in Fig. 6(a) and 6(b) for BTO and BBKT ($x = 0.15$) samples, respectively. The dotted lines in Fig. 6 indicate the energy corresponding to CBM. The optical responses start at energy level just above the bandgap for BTO in comparison to BBKT. The components of optical conductivity $\sigma_{xx}$ and $\sigma_{zz}$ have two prominent peaks at 4.12 and 8.78 eV for the BTO sample. Additionally, there are small peaks around 5.44 and 7.21 eV, as shown in Fig. 6(a). However, in the case of BBKT sample, the components of optical conductivity show peaks at 4.18 eV, and the $\sigma_{xx}$ reveals an additional plateau-like region in the incident photon energy range between 5.57 and 9.36 eV, as shown in Fig. 6(b). The components of shift current conductivity $\sigma^z_{xx}$ and $\sigma^z_{zz}$ calculated from the multi-band approach by constructing the tight-binding Hamiltonian for the BTO and BBKT samples are plotted in Fig. 6(c) and 6(d). The shift current response for both BTO and BBKT to $zz$ polarized light is positive at low energy regions but negative at high energies. Whereas the majority of the shift current response to $xx$ polarized light is positive for both systems. For BTO system, the peak of shift current conductivity $\sigma^z_{zz}$ of magnitude 6.1 μA/V$^2$ is close to the bandgap and show a plateau-like region of response in the photon energy range (5-9) eV. This is contributed by the real space charge-center shift of the valence band and conduction band, which is mainly occupied by O-2$p$ and Ti-3$d$ orbital, respectively. For BBKT, the peak in shift current conductivity $\sigma_{zz}^{z}$ is well above the bandgap with a maximum magnitude of 7.87 µA/V$^2$ at ~ 6.2 eV photon energy. The enhancement in $\sigma_{zz}^{z}$ in BBKT is originated from the additional contribution from Bi-6$p$ orbital in the conduction band.

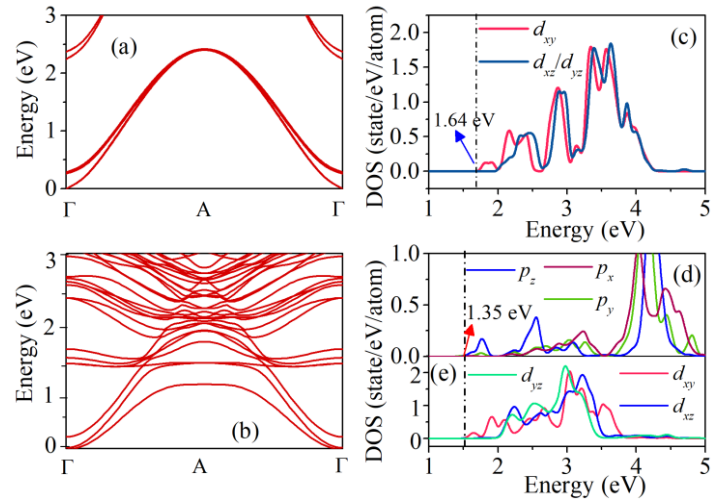

FIG. 7. Conduction band structure of (a) BTO and (b) BBKT ($x$ = 0.15) where CBM is set at zero. The corresponding orbital resolved (c) Ti-$d$ DOS for BTO, (d) Bi-$p$ and (e) Ti-$d$ DOS for BBKT sample.

In order to understand the nature of bonding and orbital specific interactions near the conduction band edge, where BPVE is expected to be largely influenced, the conduction band with CBM set at zero and partial DOS for BTO and BBKT systems are plotted in Fig. 7. Figure 7(a) and 7(b) show the dispersion in the conduction band edge states and thereby suggests the covalent nature of chemical bonding in pure BTO and BBKT.[21] To further understand the effect of the dopant on the covalence character of the BTO and BBKT systems, the Bader charge analysis is carried out on the unit cell ($x$ = 0) and supercell ($x$ = 0.15) structures. Note that, the Bader charge of the ions and its deviation from the conventional charge state provides a measure of covalency. Consequently, the charge states of the ions for corresponding supercell structures are calculated and tabulated in TABLE II. The table shows that Ti$^{4+}$ and O$^{2+}$ exhibit a significant deviation of ~1.9 $e$ and 0.8 $e$ for parent BTO systems and 1.88 $e$ and 0.79 $e$ for BBKT system. This suggests reasonable O-2$p$ – Ti-3$d$ covalent interactions in parent and doped system. Also, with Bi doping, the covalent interaction at the Ba-site increases as the charge states manifests significant deviation as large as ~1.09 $e$. On the other hand, for $Ba^{2+}$ and $K^{1+}$, the deviations are small (0.38 $e$, and 0.19 $e$), which suggests Ba and K are ionic in character.

TABLE II: Ionic charges for $Ba_{1-x}(Bi_{0.5}K_{0.5})_xTiO_3$, as estimated from the Bader charge analysis. The conventional charge state of the ions is mentioned as superscript.

| $x$ | $K^{1+}$ | $Bi^{3+}$ | $Ba^{2+}$ | $Ti^{4+}$ | $O^{2-}$ |
|---|---|---|---|---|---|
| 0 | - | - | 1.60 | 2.10 | −1.20 |
| 0.15 | 0.81 | 1.91 | 1.62 | 2.12 | −1.21 |

Figures 7(c) and 7(e) depict partial DOS of Ti-$d$ orbital for BTO and BBKT samples, where the conduction band minima is located at Γ points. The dotted lines in the partial DOS plots represent the band edges for BTO and BBKT samples. It is noteworthy to mention that upon doping, the band-edge for BBKT sample is moved towards lower energy, which could be attributed to the covalent Bi-O interaction in the sample, as reflected from the Fig 7(c)-7(e).[21] In BTO system, the CBM primarily consists of anti-bonding Ti-$d_{xy}$ state. In BBKT system, the CBM is occupied by Bi-$p_z$ state along with Ti-$d_{xy}$ state. Note that the covalent interaction between Bi-O causes the hybridization of Bi-$p$ orbitals, which lifts the degeneracy between $p_{x/y}$ and $p_z$. Consequently, Bi-$p_z$ state will be the dominant orbital character at the CBM and thereby leads to a large shift current response in $z$-direction.[21,44] Hence, the $Bi^{3+}$ doping on the $Ba^{2+}$-site in BTO system favourably alters the conduction band edge states and making it conducive for better shift current response.

IV. CONCLUSION

In summary, the lattice-site engineered BBKT systems fabricated to crystallize in single-phase tetragonal symmetry are found to exhibit enhanced ferroelectric polarization with higher $T_C$ compared to the parent BTO system. Additionally, the doped system revealed ~18% reduction in the bandgap, which is attributed to Ti-3$d$ and Bi-6$p$ orbital interaction at the conduction band through the first principle calculations. The observed linear and sinusoidal current response with intensity and light polarizer angle, respectively, strongly suggests that the PV responses of BBKT samples studied through the phenomenological approach are the consequence of BPVE. Importantly, $x = 0.125$ composition displays the highest $V_{OC}$, which is ~8 times higher than BTO and comparable to reported results on other polycrystalline systems. The difference in observed $J_{SC}$ between the parent and doped system is correlated to the consequence of the difference in $\beta_{31}$ value. The theoretical studies on the PV effect revealed a large shift current response in BBKT (7.87 µA/V$^2$) over BTO (6.1 µA/V$^2$). This is correlated to the real space shift of the O-2$p$ occupied valence electron and Bi-6$p$ occupied conduction electron, in addition to the charge-center shift that occurred in the BTO system. The degenerate Bi-$p_z$ state near CBM makes the BBKT sample conducive to the large shift current response in the $z$-direction. Overall, the combined PV study of the modified BTO system through phenomenological and shift current theory is one step towards understanding the bulk photovoltaic phenomenon.

## SUPPLEMENTARY MATERIAL

The action spectrum measured on $x = 0.125$ BBKT sample under the Xenon Arc lamp illumination using the monochromatic filters at 100 mW/cm$^2$ light intensity is shown in Fig. S1.

## ACKNOWLEDGMENTS

The authors acknowledge Prof. M. S. Ramachandra Rao and Dr. M. Rath for P-E measurements. B S would like to thank Prof. Jeroen van den Brink for FPLO code packages and IFW cluster facilities.

## DATA AVALABILITY STATEMENT

The data that support the findings of this study are available within the article.

**Figure Captions:**

FIG. 1. (a) The XRD patterns of $x$ = 0.0, 0.05, 0.075, 0.1, 0.125 and 0.15 BBKT samples. (b) Enlarged XRD pattern around (002)/(200) diffraction peaks. The solid lines are drawn for the reference with respect to the parent sample.

FIG. 2. XRD patterns represent Bragg profiles of (002) and (200) for the BBKT system fitted *P4mm*, *P4mm+Amm*2, and *P4mm+Pm* phase models for (a) $x = 0.05$, (b) $x = 0.075$, (c) $x = 0.10$, (d) $x = 0.125$, and (e) $x = 0.15$ samples. Open symbol, solid line, and vertical line represent experimental points, refined data, and Bragg profiles, respectively. The bottom solid lines represent the error between experimental and calculated data points.

FIG. 3. (a) P-E hysteresis loops and (b) temperature-dependent real parts of the permittivity ($\varepsilon_1$) for $x = 0.0, 0.05, 0.075, 0.10, 0.125$, and $0.15$ samples.

FIG. 4. (a) The K-M plots of BBKT samples and (b) the corresponding bandgap variation with compositions. (c) Super cell structure and band structure of BTO. (d) Super cell structure and band structure of BBKT. DOS of (e) BTO-unit cell and (f) BBKT-223 super cell.

FIG. 5. (a) Schematic of experimental set-up used for PV measurements. (b) *J*-V characteristic under 405 nm illumination measured at 11.9 mW/mm$^2$ light intensity for $x = 0.05$ to $0.15$ samples. The inset shows the corresponding plot for $x = 0$ sample. The $J_{SC}$ plotted as a function of (c) light intensities and (d) light polarization angle for BBKT samples. The inset shows the schematic of the measurement geometry. (e) Extracted $\beta_{31}$ for $x = 0.0, 0.05, 0.075, 0.10, 0.125$, and $0.15$ samples. (f) The composition-dependent experimentally observed $V_{OC}$ and calculated $v_{ph}$ from the phenomenological model.

FIG. 6. The optical conductivities $\sigma_{xx}$ and $\sigma_{zz}$ plotted for (a) BTO and (b) BBKT. The shift current conductivities $\sigma_{xx}^z$ and $\sigma_{ZZ}^Z$ of (c) BTO and (d) BBKT ($x = 0.15$) system.

FIG. 7. Conduction band structure of (a) BTO and (b) BBKT ($x = 0.15$) where CBM is set at zero. The corresponding orbital resolved (c) Ti-*d* DOS for BTO, (d) Bi-*p* and (e) Ti-*d* DOS for BBKT sample.

**Table Captions**:

TABLE I. The refinement parameters for $x$ = 0.0, 0.05, 0.075, 0.10, 0.125, and 0.15 BBKT samples.

TABLE II: Ionic charges for $Ba_{1-x}(Bi_{0.5}K_{0.5})_xTiO_3$, as estimated from the Bader charge analysis. The conventional charge state of the ions is mentioned as superscript.

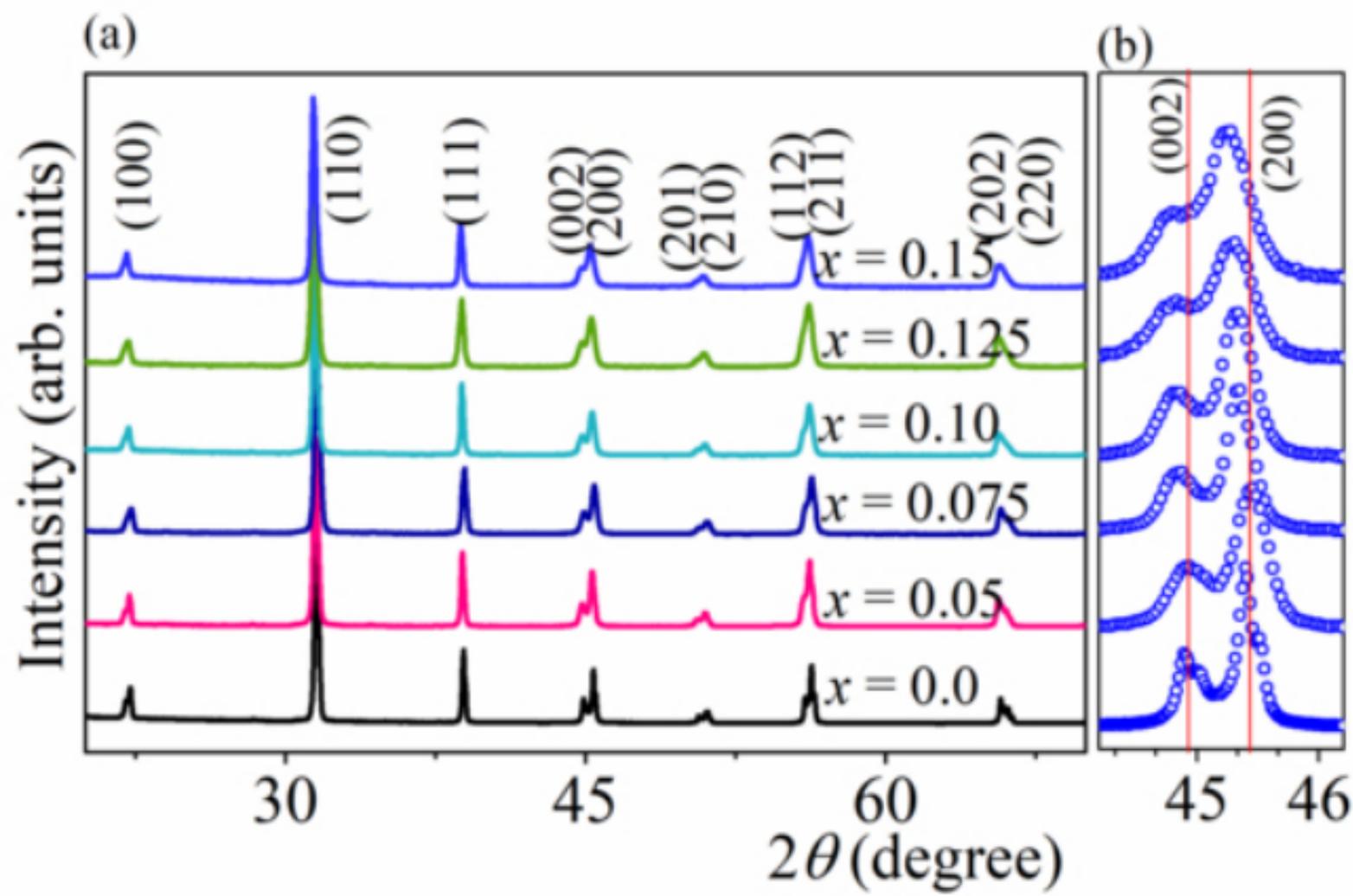

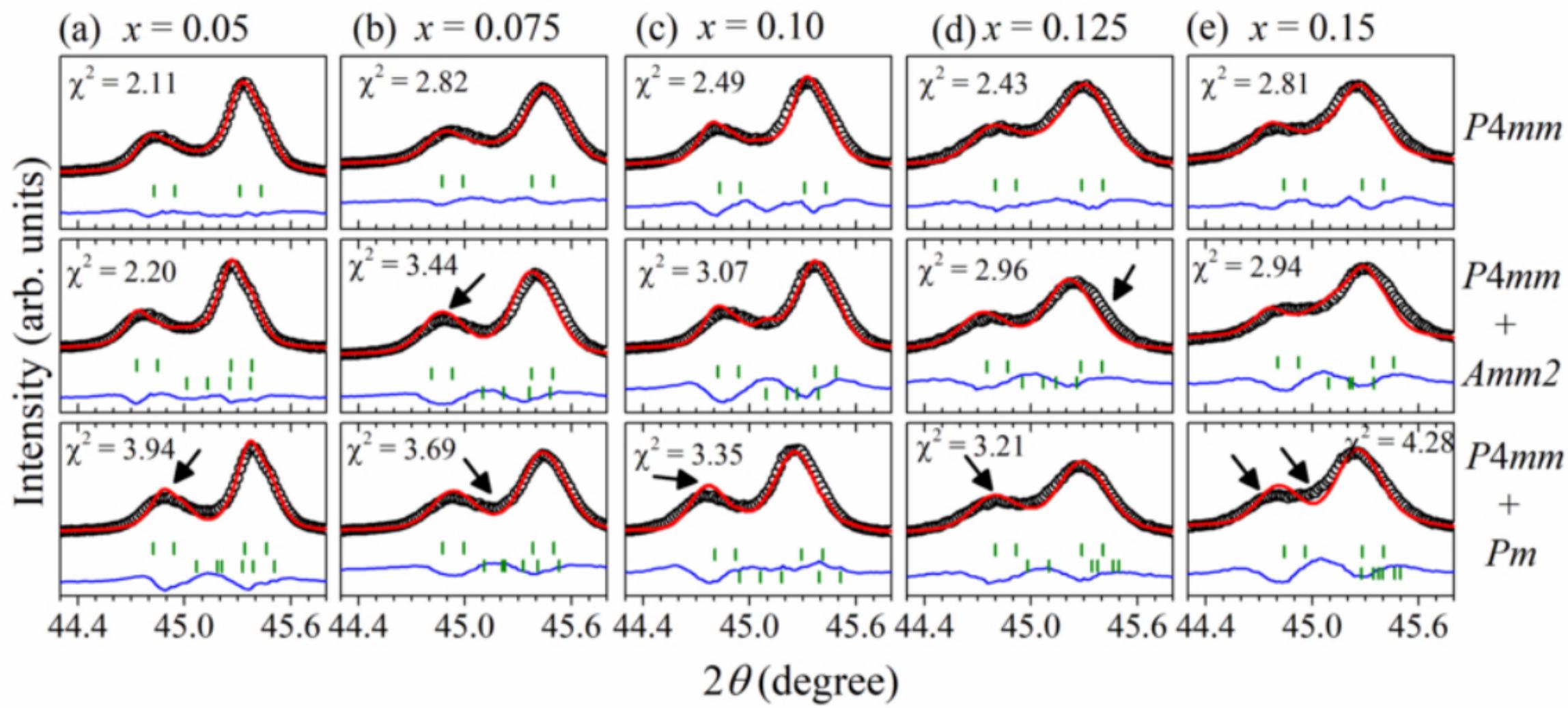

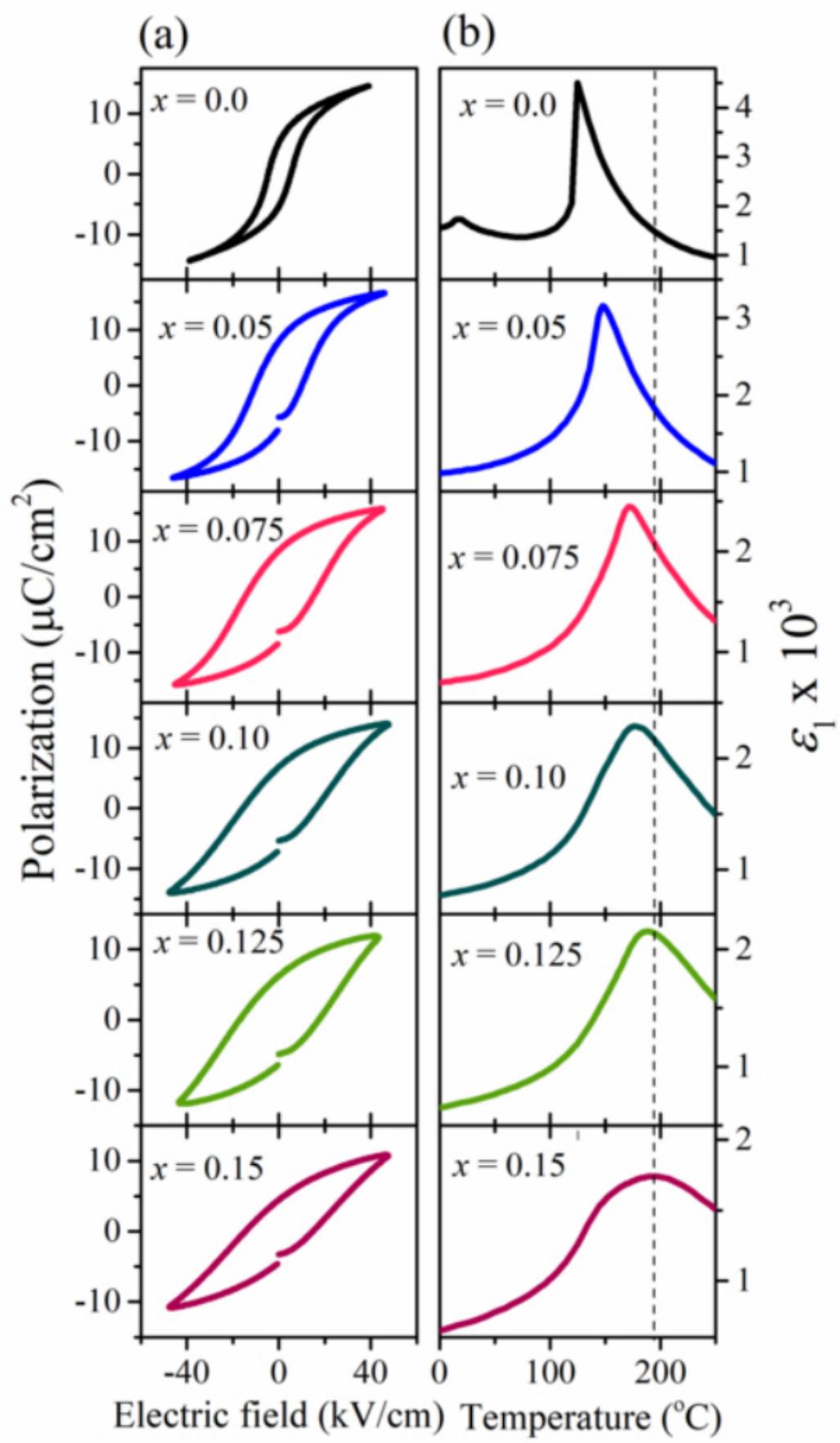

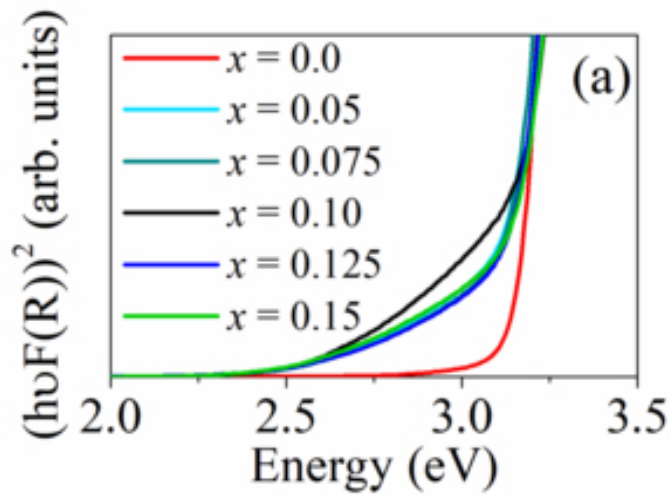
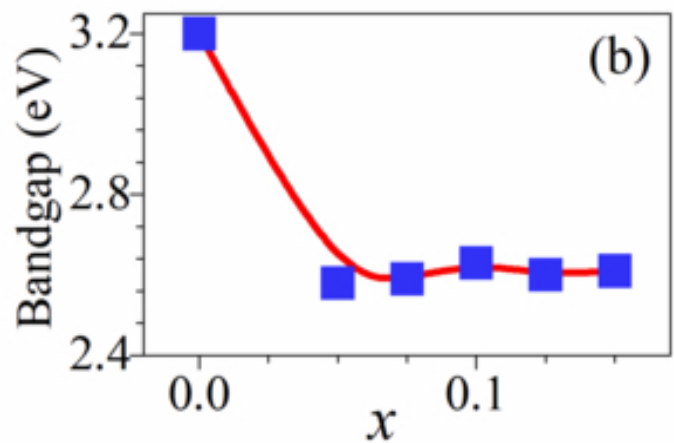
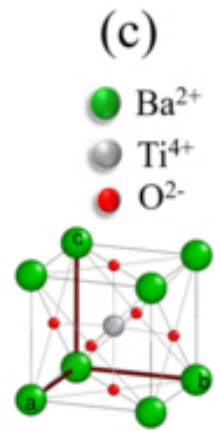
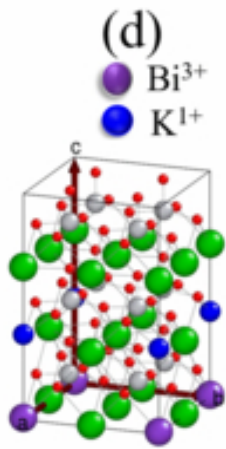
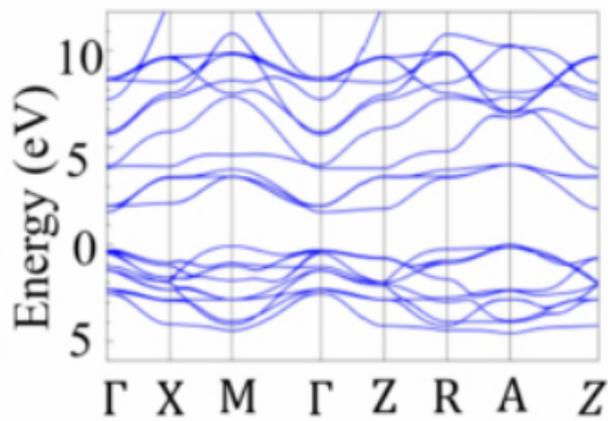
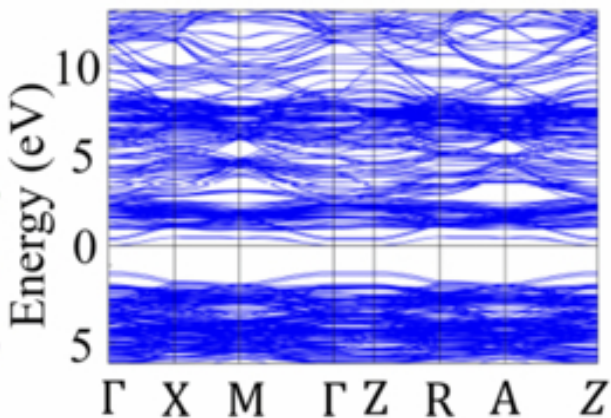
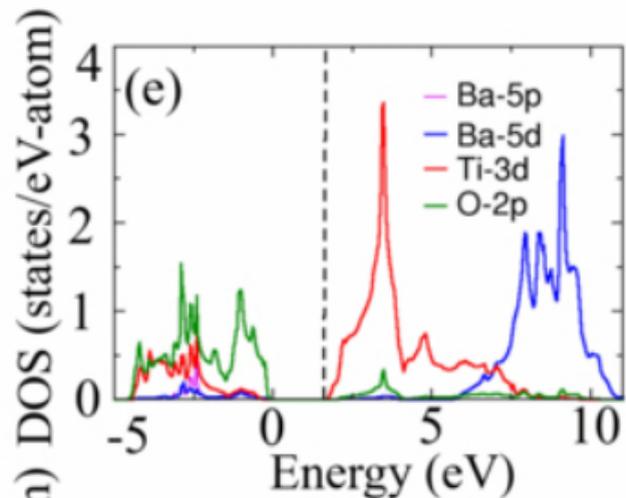
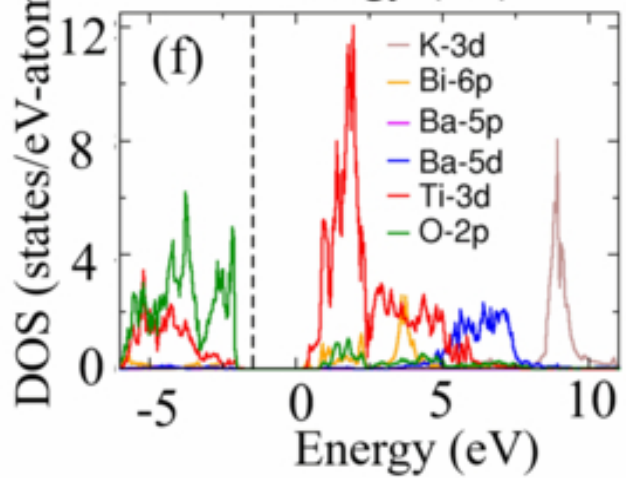

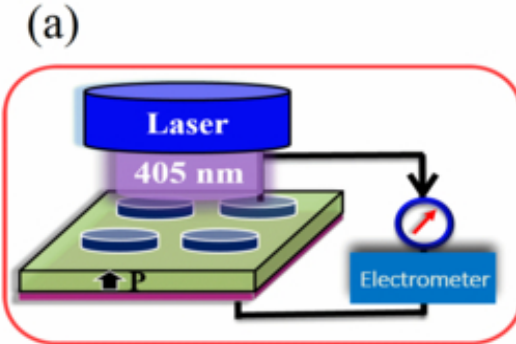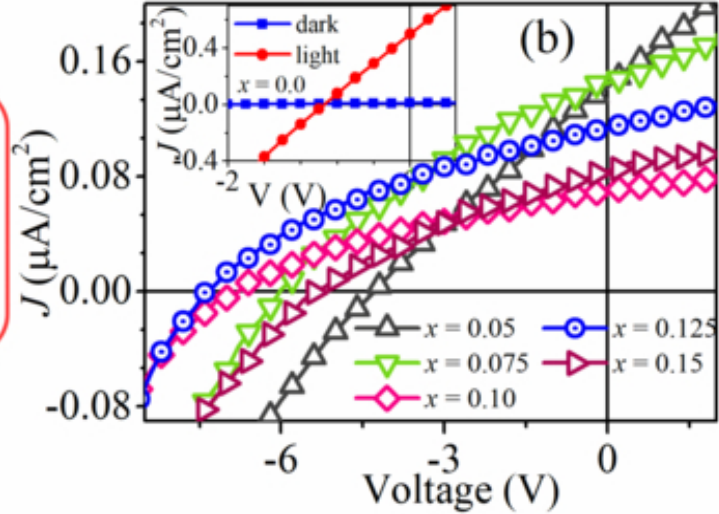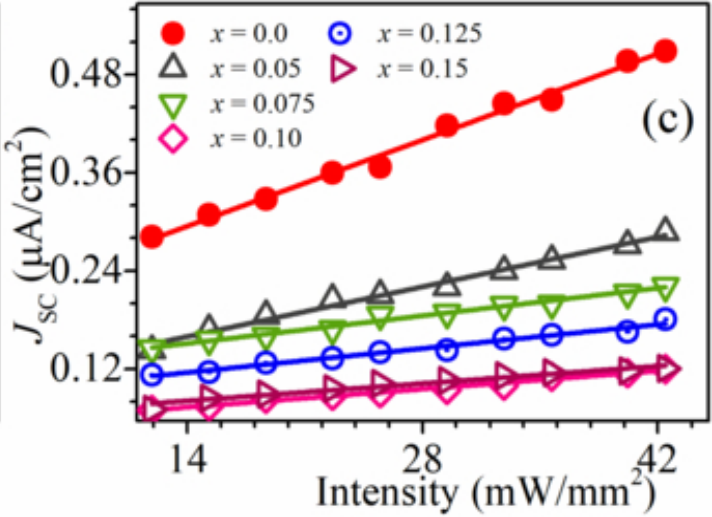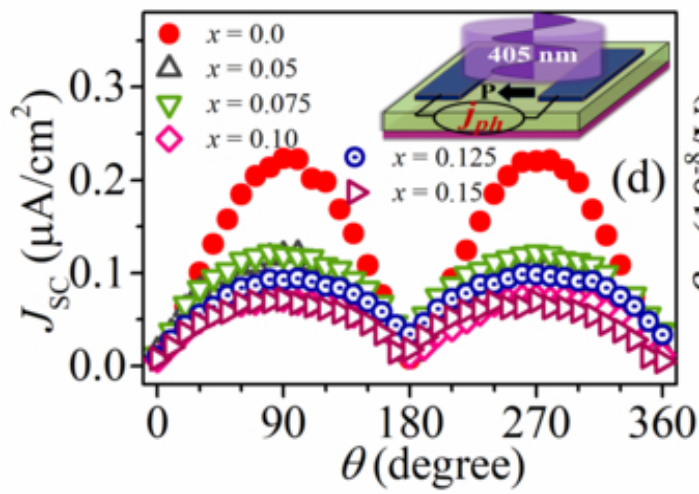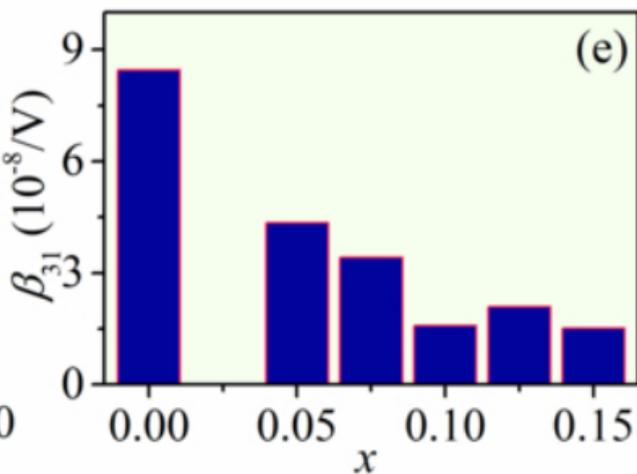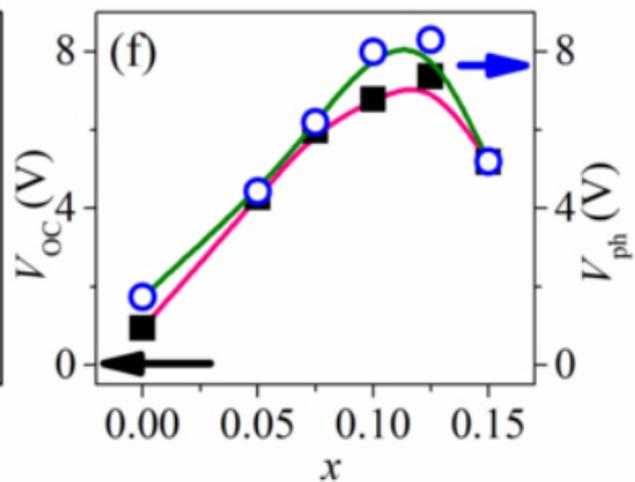

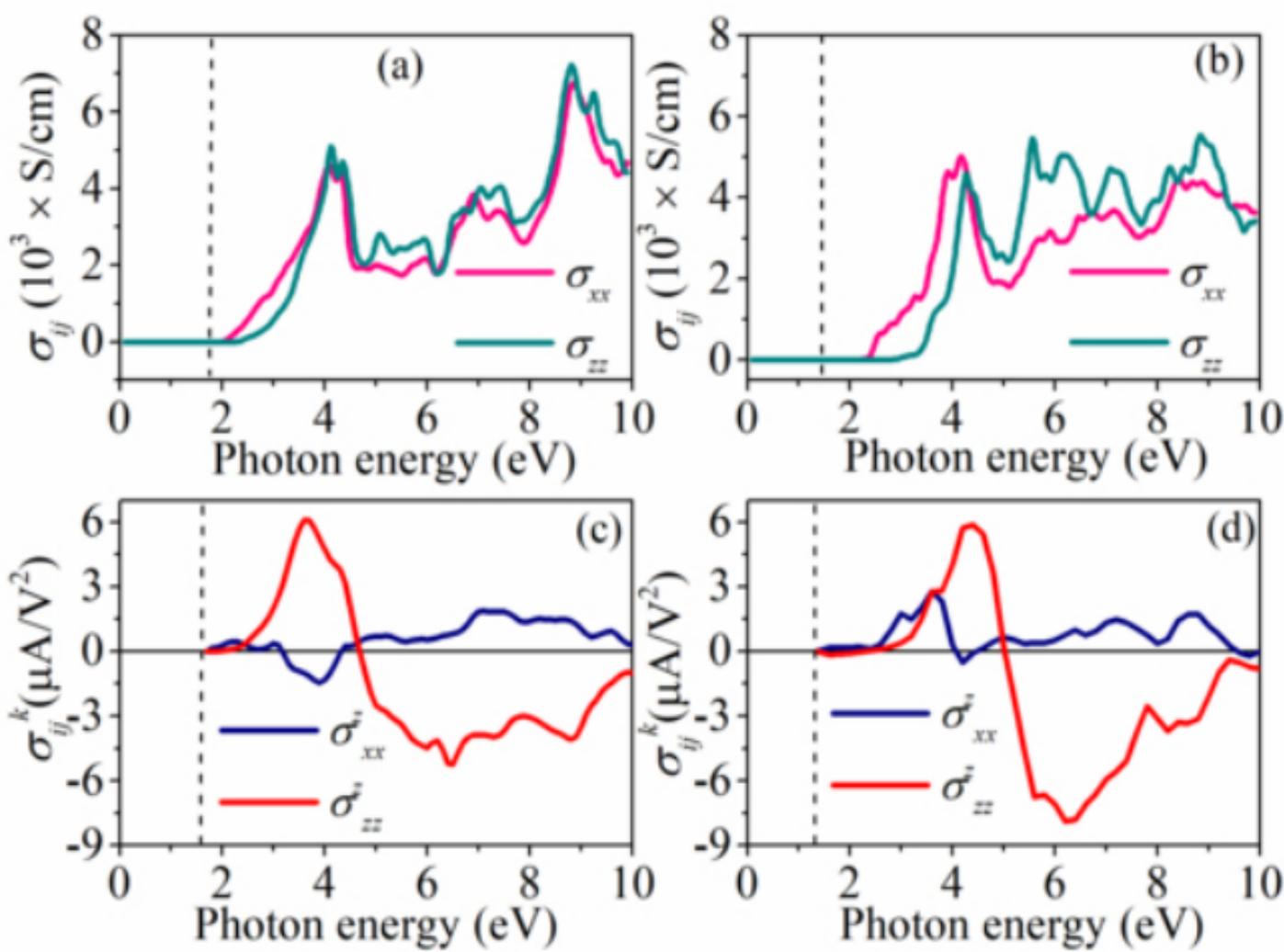

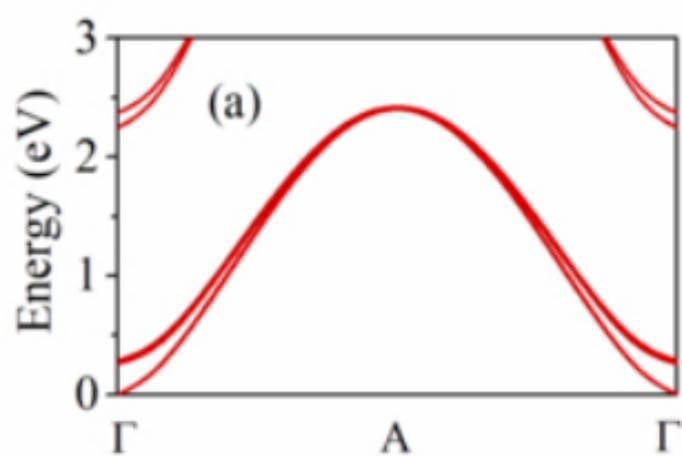
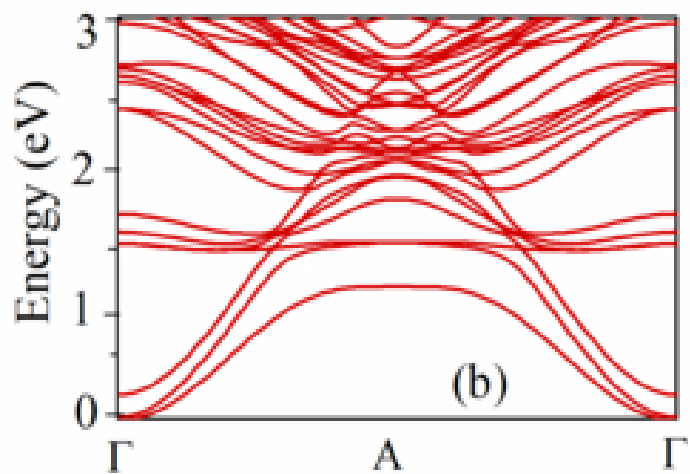
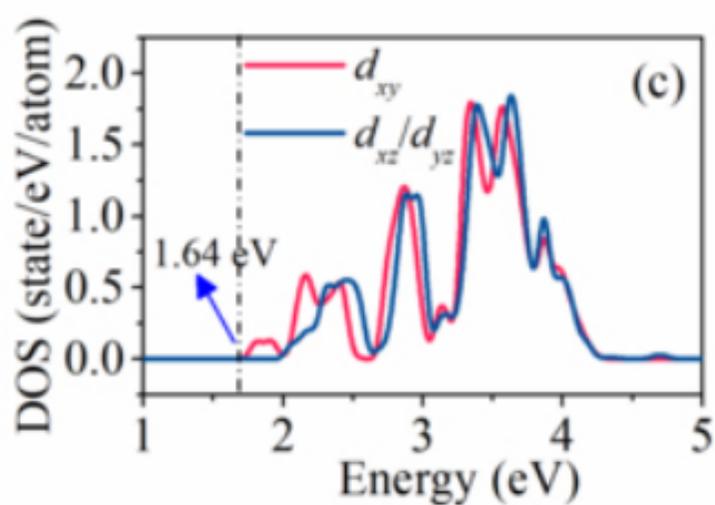
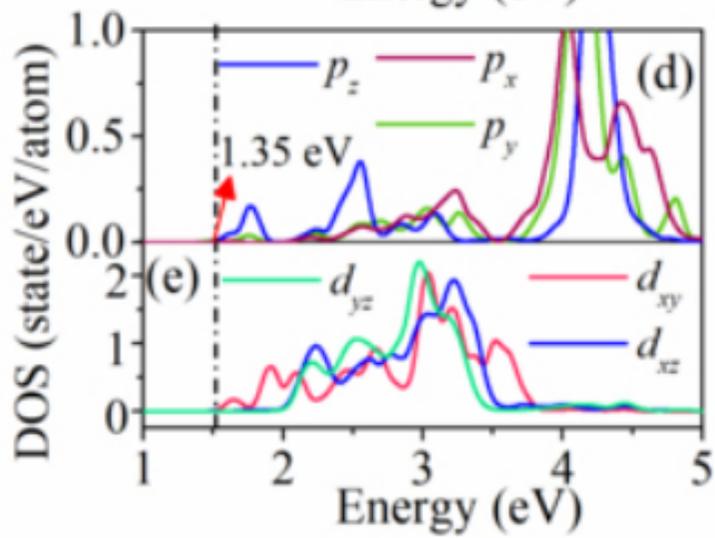